\documentstyle[aps,epsf,floats]{revtex}
\begin{document}

\begin{flushright}

JLAB-THY-04-35\\
October 20, 2004 \\
JINR P2-10717 \\
June 14,1977\\
\end{flushright}

\vspace{1cm}

\begin{center}
{\Large \bf DEEP ELASTIC PROCESSES OF  \\
\vspace{0.3cm}
COMPOSITE PARTICLES IN FIELD THEORY\\ 
\vspace{0.5cm} AND ASYMPTOTIC FREEDOM$^*$}

\end{center}

\vspace{1cm}

\centerline{\Large \bf A.V. Radyushkin$^{**}$}

\begin{center}
{\em $^*$The  investigation has been performed 
(and completed in June 1977) 
at the  } \\ 
{\em  Laboratory of Theoretical  Physics, 
JINR, Dubna, Russian Federation}

 \vspace{5mm}

{\em English translation and comments: October 2004} 

\vspace{1cm}

{\em $^{**}$Present address: 
Physics Department, Old Dominion University,
Norfolk, VA 23529, USA} \\ {\em and} \\{\em  Theory Group, Jefferson Lab,
Newport News, VA 23606, USA}

\end{center} 

\begin{abstract}

\noindent 
This is an English   translation of my 1977 Russian preprint.
It contains  the first explicit definition of the pion 
distribution amplitude (DA),  the 
expression for the pion form factor asymptotics 
in terms of the pion DA,  and  formulates  the pQCD parton 
picture for hard exclusive processes. 


\noindent Abstract of the original paper: \\
The large $Q^2$ behavior of the pion electromagnetic form factor
is explicitly calculated in the 
non-Abelian gauge theory to demonstrate a field-theoretical
approach to the deep elastic processes of composite particles.
The approach is equivalent to a new type 
of parton model.   

\end{abstract}

\newpage

At present {\bf \{1\}}, it is usually accepted that the hadrons are complex  objects
composed  of quarks interacting through a mediating gluonic field {\bf \{2\}}. 
The approximate scaling in  deep inelastic 
scattering indicates that the effective coupling of the 
quark-gluon interaction 
is small in the region of large transferred momenta. 
Hence, there is some justification for the 
 hope that the quantum-field calculations 
based on the use of perturbation theory can be appied 
in the region of large momentum transfers.  
An efficient technique based on the use of the operator expansions
and the renormalization group (RG) methods   
was applied to the study of the deep inelastic scattering processes
\cite{Politzer:fr}. Incorporating  the 
recent studies of the  pion form factor behavior 
at large $Q^2$ with the help of the operator expansions 
\cite{Goldberger:vp} and the analysis of the Feynman graphs
in the $\alpha$-representation \cite{Efremov:1976np},
we investigate deep elastic processes involving composite  
particles with the help of  a technique analogous to that
used in \cite{Politzer:fr}. 

Consider the pion electromagnetic form factor,
treating the pion as a bound state of spin 1/2 quarks. 
The form factor of such a system 
can be written in terms
of the Bethe-Salpeter wave function $\chi$ as shown in 
{\it Fig. 1a}. Due to  property
$\chi = K \otimes \chi$, where $\chi$ is the Bethe-Salpeter 
kernel, we can rewrite $F_{\pi}$ as shown in {\it Fig. 1b}.
  \epsfxsize=14cm
 \epsfysize=7cm
 \hspace{4cm}  \epsffile{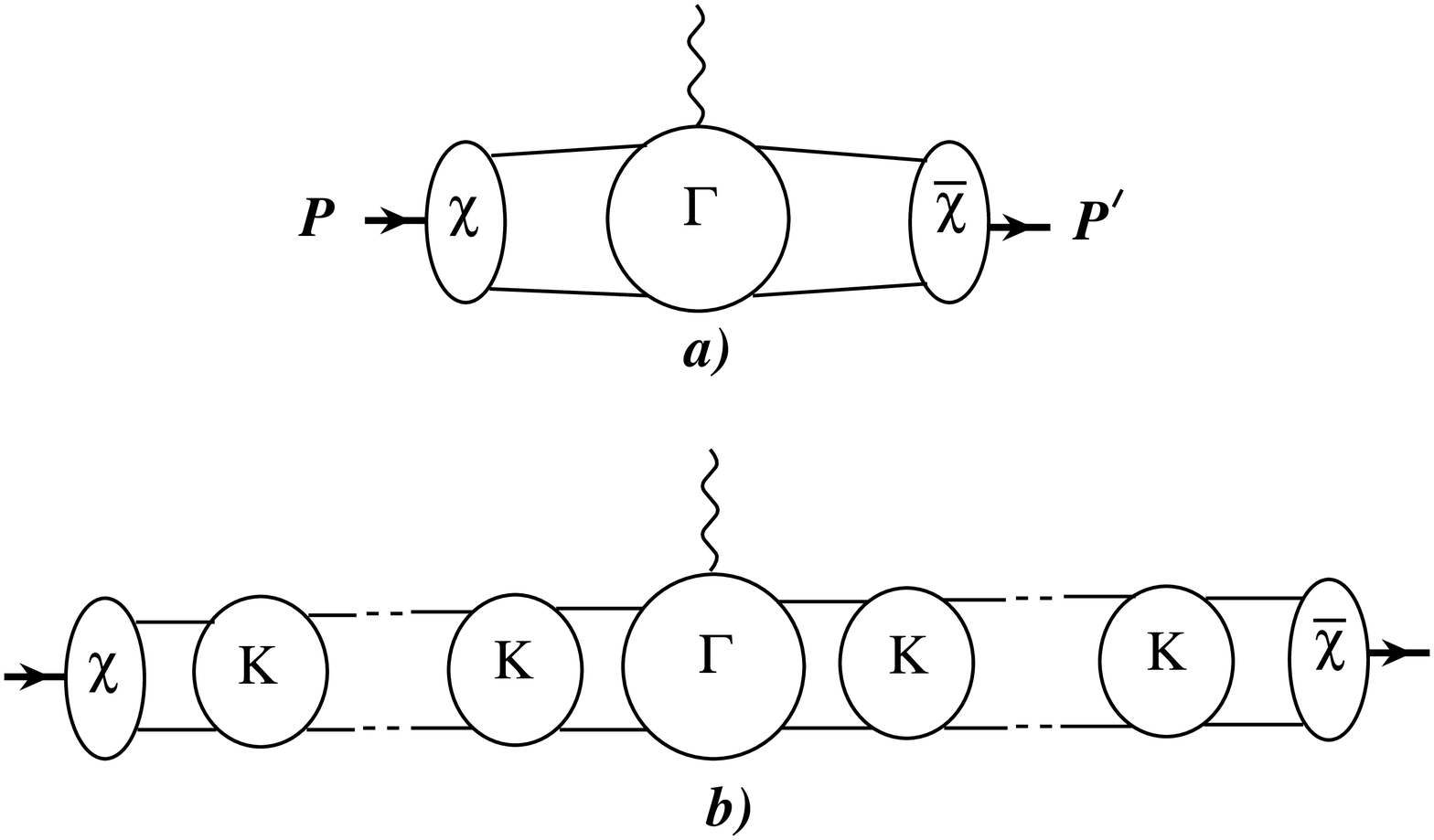}
 \begin{center}  
   {Fig. 1. Electromagnetic form factor of a composite pion.}
   \end{center}
   
   \vspace{5mm}
   
Each graph  can be formally written 
in the   $\alpha$-representation (see, e.g., \cite{Bogolyubov:nc},
\cite{Efremov:vx}): 
\begin{eqnarray}
R^{\mu} (P,P') = g^2 \left (\frac{g^2}{16\pi^2} \right )^{N_{\rm loop}}
\int_0^{\infty} \frac{\prod d\alpha_{\sigma}}{D^2(\alpha)} \, 
G^{\mu}(\alpha, P, P')\, \exp\left \{ iq^2 
\frac{A(\alpha)}{{ D}(\alpha)} +i I(\alpha, m_{\pi}^2) \right \} \ ,
\label{1}  
\end{eqnarray}
where the functions $ D, G, A$ and $I$ are determined by the 
topological structure of the diagram. Note that  $\chi, \bar \chi$ 
vertices should be treated as subgraphs rather than points.
One of the  most important results  of the analysis of the
Feynman graphs in the $\alpha$-representation is that the large-$Q^2$
behavior in a theory with 
dimensionless coupling constant is determined 
by integration over a region of the $\alpha$-parameters 
belonging to subgraphs the contraction of which  into 
point eliminates the dependence of the graph on the large 
variable $Q^2$. For the pion electromagnetic form factor,
integration over small $\lambda_V$ ($\lambda_V = \sum \limits_{\sigma \in V} 
\alpha_{\sigma} $)  gives the leading contribution 
  $F_V^{\rm lead} (Q^2) \sim Q^{-\Sigma \, t_i +2}$, where $t_i$
  is the twist ($ \equiv$ dimension minus spin) of the field 
  describing the $i$-th   external line of the subgraph $V$, 
  and $Q= \sqrt{-q^2}$.  Taking into account that, in a 
  theory with vector gluons, $t=1$ for  quark fields, but $t=0$ for 
  vector  fields (both for  gluons and the photon),  one 
  may conclude that the leading contribution can be 
  given only by subgraphs having 4 external quark lines
  and an arbitrary number 
  of  gluon lines ({\it Fig. 2}).

  Using the Mellin representation 
   \begin{equation}
  F_{\pi} (Q^2) = \frac1{2\pi i} \int_{-i\infty}^{ i\infty} 
  \Phi (J) (Q^2)^J dJ \ , 
  \label{2}
  \end{equation} 
  one can   say   that  integration over the region 
  $0 \leq \lambda_V \leq 1/\mu^2$ gives a pole 
  $(J+1)^{-1} (1/\mu^2)^{J+1} \bar f \otimes E \otimes f$,
  where $E$ is the contribution due to the subgraph $V$ 
  and $\bar f, f$ are the contributions  of 
  the left and right weakly connected parts 
  resulting from the contraction of $V$ into point.
  
   \begin{center}
  \epsfxsize=10cm
 \epsfysize=3cm
  \hspace{1cm}  \epsffile{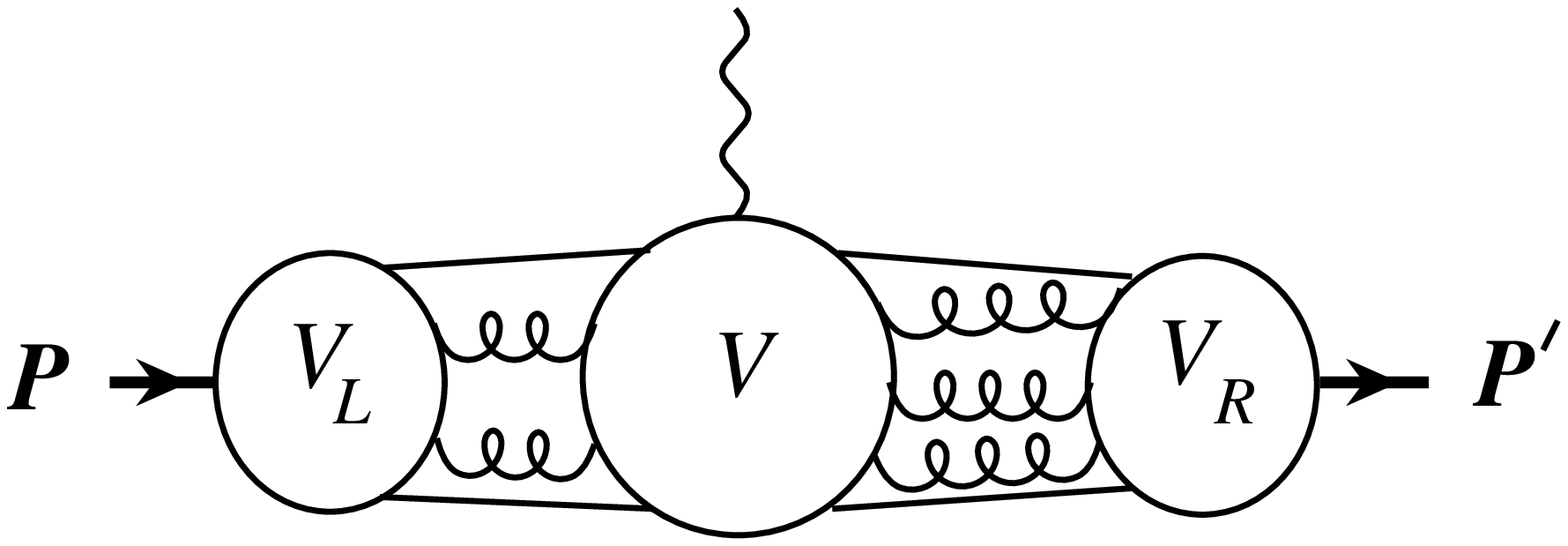}
   \end{center}
 \centerline{Fig. 2. Subgraph producing the leading large-$Q^2$ contribution.}  
  
   \vspace{5mm}
  
  The functions $f, \bar f$ do not depend on $J$ 
  because the contraction of $V$ into point 
  eliminates the dependence on $Q^2$. 
  Summation over all possible subgraphs looks like
  a perturbative series, with $E(Q^2/\mu^2, g(\mu^2))$
  describing short-distance interactions while $f$ and $ \bar f$ 
   correspond to the properly normalized 
   quark-pion vertices:
  \begin{eqnarray}
  F_{\pi} (Q^2) = \frac1{Q^2} \sum_{n=0}^{\infty}\sum_{m=0}^{\infty}
  \bar f_m^I(\mu^2,g, P') E_{mn}^{IK} (Q^2/\mu^2, g) f_n^K(\mu^2,g, P) \ ,
  \label{3}
   \end{eqnarray}
    where $I,K$ denote the spinor indices.
   The leading contribution can be obtained 
   only when $f$ and $ \bar f$  are projected on the axial
   structure \cite{Goldberger:vp,Efremov:1976np,Polyakov:fx}.
   The functions $f_n^A \equiv a_n$ are defined through
   matrix elements of the twist-2 operators
   $\bar \psi \gamma_5 \gamma_{\mu_1}
   \stackrel{\leftrightarrow}{D}_{\mu_2}  \ldots 
    \stackrel{\leftrightarrow}{D}_{\mu_{n+1}} \psi$ :
\begin{eqnarray}
2^n i^{n-1} \langle 0 | \bar \psi (0) \gamma_5 \{ \gamma_{\mu_1}
   \stackrel{\leftrightarrow}{D}_{\mu_2}  \ldots 
    \stackrel{\leftrightarrow}{D}_{\mu_{n+1}} \} \psi (0) | P \rangle
    =  \{ P_{\mu_1}\ldots P_{\mu_{n+1}} \}  \, a_n( \mu^2, g(\mu^2))\, 
    \frac{1+(-1)^n}{2} \  , 
\label{4}
   \end{eqnarray}
 where $\stackrel{\leftrightarrow}{D}_{\mu_{i}}$ 
 is the covariant   derivative acting on the quark fields,
 $\{ \ \ \}$ means that the corresponding function is symmetric 
 in $ \mu_1, \dots, \mu_{n+1}$ and all its contractions 
 with $g^{\mu_i \mu_j}$ are zero, and $\mu^2$ serves as the renormalization
 parameter. 
  
  Applying  $\mu\,d / d\mu$ to both sides of Eq.(\ref{4}), we obtain 
  a renormalization group equation
  \begin{eqnarray} 
  \sum_{n=n'}^{\infty}\sum_{m=m'}^{\infty}
  \left \{ \left [ \mu \, \frac{\partial}{\partial \mu}+  \beta (g)
  \frac{\partial}{\partial g} \right ] \, \delta_{m'm}\delta_{nn'}
  + \bar z_{m'm} \delta_{nn'} + \delta_{m'm}z_{nn'} \right \} E_{mn} = 0 \ , 
  \label{5}
   \end{eqnarray}
 where
  \begin{eqnarray} 
  \left [ \mu \, \frac{\partial}{\partial \mu}+  \beta (g)
  \frac{\partial}{\partial g} \right ] a_n ( \mu^2, g(\mu^2))=
  \sum_{n'=0}^{n}z_{nn'}a_{n'} \ . 
     \label{6}
   \end{eqnarray}
 According to (\ref{6}), there is  mixing between  operators
 with different spin (but the same twist).

 Let us illustrate the method of calculations on the example of the
 lowest-order diagram ({\it Fig. 3a)}.
 The $\alpha$-representation for the leading contribution of such a diagram 
 in a non-Abelian gauge theory has the following structure:
   \begin{eqnarray} 
  g^2 C_2 (R) \sum_{V_L,V_R} \left (\frac{g^2}{16\pi^2} \right )^{N_L+N_R}
& &\int \frac{\prod d\alpha^{(L)}}{D^2 (V_L)} 
 \,  e^{iI(\alpha_{\sigma}^{(L)},m_{\pi}^2)} \, 
  \frac14 \,{\rm Sp} \{\gamma_5 \gamma_{\nu} G_L\}
  \int \frac{\prod d\alpha_{\sigma}^{(R)}}{D^2 (V_R)} \,
  e^{iI(\alpha^{(R)},m_{\pi}^2)} \,
  \frac14 \, {\rm Sp} \, \{\gamma_5 \gamma_{\lambda} G_R\}
  \nonumber \\ & & \int d \alpha d \beta 
  \, {\rm Sp} \, \{\gamma_5 \gamma^{\nu}\gamma^{\alpha} 
  \gamma_5 \gamma^{\lambda}\gamma_{\alpha} \hat P \gamma^{\mu} \} \, 
  \frac{L(\alpha^{(L)}) D(V_R)}{D} \, e^{i\{\alpha \frac{LD(V_R)}{D}+
  \beta \frac{LR}{D}\}q^2 + O(\lambda^2)} \ ,
  \label{7}
   \end{eqnarray}
where $ \alpha_{\sigma}^{(L)}\in V_L,  \alpha_{\sigma}^{(R)}\in V_R$.
We also have taken into account that,  for  the leading contribution,
the function $G_L$ ($G_R$) depends only on $P$ ($P'$)
and that $D=D(V_L)D(V_R)+ O(\lambda)$, where $\lambda = \alpha +\beta$.
The meaning of the 2-trees $L,L',R,R'$ is clear from {\it Fig. 3b}.

 \begin{center}
  \epsfxsize=16cm
 \epsfysize=5cm
   \epsffile{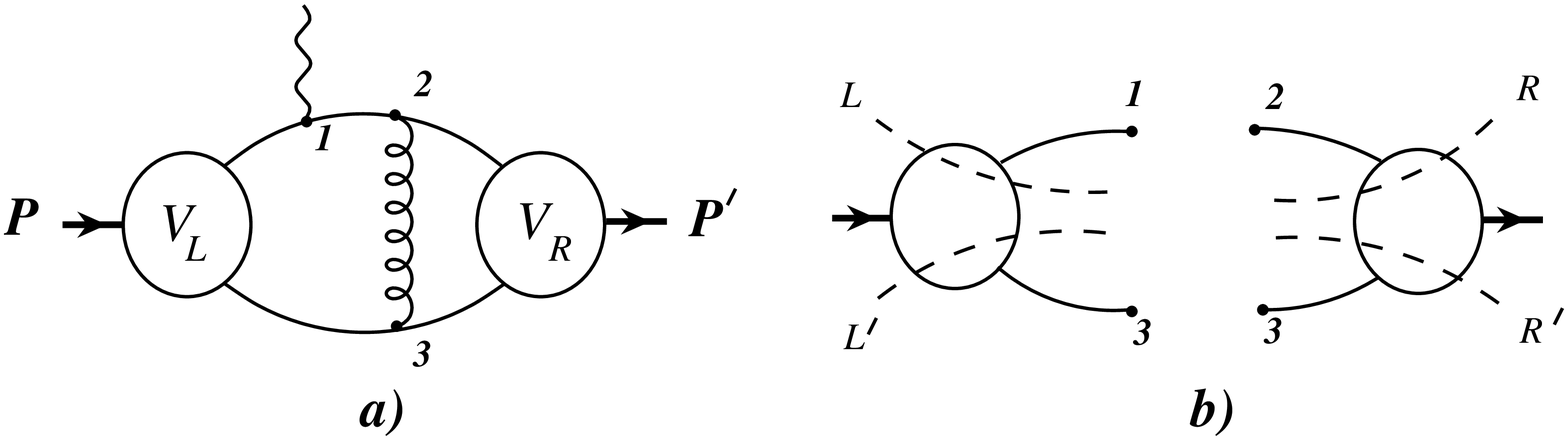}
   {Fig. 3. Structure of the lowest-order diagram.}
   \end{center}
    
    \vspace{5mm}
  
Integration in the neighboorhod of $\alpha \sim 0, \ \beta \sim 0$ gives
  \begin{eqnarray} 
  g^2 C_2 (R) \, \frac{2P^{\mu}}{q^2}
   \sum_{V_L,V_R} \left (\frac{g^2}{16\pi^2} \right )^{N_L+N_R}
& &\int \frac{\prod d\alpha^{(L)}}{D^2 (V_L)} 
 \,  \frac{e^{iI(\alpha_{\sigma}^{(L)},m_{\pi}^2)}}{1-L_-/D(V_L)} 
  \, G^A(V_L) 
  \int \frac{\prod d\alpha_{\sigma}^{(R)}}{D^2 (V_R)} 
 \, \frac{ e^{iI(\alpha^{(R)},m_{\pi}^2)} }  
 {1-R_-/D(V_R)} \, G^A(V_R)
   \ . 
  \label{8}
   \end{eqnarray}
We used the fact that for the diagrams of {\it Fig. 3b}  
type $R_+ = D(V_R)$, where $R_{\pm}=R'\pm R$, and hence 
$  R /D(V_R)= (1-R_-/D(V_R))/2$. Expanding into series
$1/(1-R_-/D(V_R)) = \sum (R_-/D)^n$ we obtain an expression of
Eq. (3) type. The factor $(R_-/D)^n$ just means that 
the corresponding operator contains $n$ derivatives.
Due to  symmetry between $R$ and $R'$, the contributions
corresponding to odd $n$ vanish. Let us introduce
the partonic wave function 
 \begin{eqnarray} 
 \widetilde \varphi (\xi,\mu^2) 
 ={\rm Reg}_{{\mu^2}} 
   \sum_{\rm diagr} \left (\frac{g^2}{16\pi^2} \right )^{N_d}
\int \frac{\prod d\alpha_{\sigma}}{D^2 (\alpha)}\, G^A(\alpha,P)\,
 e^{iI(\alpha,m_{\pi}^2)} \, \delta(\xi -R_-/D) 
  \ . 
  \label{9}
   \end{eqnarray}
   Since $|R_-/D| \leq 1$, the function vanishes for $|\xi|\geq 1$. 
 
  In other  words, the function   $\widetilde \varphi (\xi,\mu^2) $
  is such a function, the $n$-th moment of which is equal 
  to $a_n (\mu^2)$ {\bf  \{3\}}:
  \begin{eqnarray}  
  \int_{-1}^1 d\xi \, \xi^n \widetilde \varphi (\xi,\mu^2) 
  =a_n (\mu^2)\, \frac{1+(-1)^n}{2}\ .
      \label{10}
   \end{eqnarray}  
Hence, $  \widetilde \varphi (\xi) =    \widetilde \varphi (-\xi)$.

The function $\widetilde \varphi$ can be normalized by noticing that
the magnitude of the matrix element of the axial current
is known from the $\pi \to \mu \nu$ process:
 \begin{eqnarray} 
 \langle 0 | \bar \psi (0) \gamma_5 \gamma_{\mu} \psi (0)|P\rangle
 =i \sqrt{2} f_{\pi} P_{\mu} \ ,
   \label{11}
   \end{eqnarray}  
 with $f_{\pi} \approx 100$\,MeV. Since the axial current
 has zero anomalous dimension, the relation (11) 
 is valid for any $\mu^2$.  That is why we take
 $\widetilde \varphi (\xi,\mu^2) = \frac{\sqrt{2}}{2} f_{\pi} 
 a(\xi, \mu^2)$. The final result then has the following form 
  \begin{eqnarray} 
 F_{\pi}^{\rm lead} (Q^2) = 16 \pi \alpha_s (Q^2) \, \frac{f_{\pi}^2}{Q^2}
 \, \frac{C_2(R)}{N_c} \, |\gamma (Q^2)|^2 \ ,
   \label{12}
   \end{eqnarray}  
    where
  $$
    \gamma (Q^2) = \int_0^1 a(\xi, Q^2)\frac{d\xi}{1-\xi^2}  \  \   \  \ ,
    \  \  \  \  \left | \int_0^1 a(\xi, Q^2) d\xi \right | =1  \  .
   $$
 
 The representation (3) is equivalent to the following parton picture
 for  deep elastic scattering.
 The  splitting of the pion into two quarks with momenta
 $x_1P, x_2 P$ is described by the wave function 
 $\varphi (x_1, x_2,\mu^2)\, \delta(1-x_1-x_2)$, while 
   the fusion of the quarks with momenta $y_1P',y_2P'$  
  into the  pion is described by the wave function 
   $\varphi^* (y_1, y_2,\mu^2)\, \delta(1-y_1-y_2)$.
   The  amplitude for $x_1P, x_2 P \to y_1P',y_2P'$ 
   transition is constructed according  to the usual rules:
    \begin{eqnarray} 
 F_{\pi} (Q^2) = \int_0^1 dx \, dy \, \varphi^* (y,1- y,\mu^2)\,
 E(x,y,P,P',\mu^2) \, \varphi (x, 1-x,\mu^2)  \  .
  \label{13}
   \end{eqnarray}  
  For the diagram {\it 3a} {\bf \{4\}}
  $$
  P_\mu E = \frac1{xyq^2} \,  \frac1{xq^2} \, 
  {\rm Sp} \left \{ \gamma^{\mu} \frac{\gamma_5 \hat P}{4}\gamma^{\alpha}
    \frac{\gamma_5 \hat P'}{4} \gamma_{\alpha} (\hat P'-xP)\right \} \ ,
  $$
 the wave functions are symmetric  $\varphi (x_1, x_2) = \varphi (x_2, x_1)$
 and $\widetilde  \varphi (\xi) =  
 \varphi \left (\frac{1+\xi}{2},\frac{1-\xi}{2} \right )$. 
 This physical interpretation suggests that $\widetilde  \varphi (\xi,\mu^2)$
 is maximal for $\xi=0$ and vanishes  for $\xi=1$.
 Taking for simplicity $|a(\xi, \mu^2)|=\frac32 (1-\xi^2)$ {\bf \{5\}}
 (where 3/2 is determined by the normalization condition) and incorporating 
 that the analysis of scaling violation in deep inelastic 
 scattering under the assumption that the asymptotic 
 freedom takes place, i.e., that 
 $\alpha_s (Q^2) =\frac{4\pi}{9 \ln (Q^2/\Lambda^2)}$,
 gives $\Lambda =0.5$\,GeV \cite{Parisi:1976fz,DeRujula:1976tz,Gluck:1976ew}, 
 we obtain that $\alpha_s (Q^2=2\,$GeV$^2) =0.7$,
 and, taking into account that $C_2(R)=4/3, N_c=3$, we find 
 $F_\pi (Q^2=2) = 0.18$. The formula $F_\pi (Q^2) =(1+Q^2/0.47)^{-1}$ (the 
 extrapolation of  the $\rho$-meson pole well desribing experimental data)
 gives $F_\pi (Q^2=2) = 0.19$. The factor ${m_\pi^2}/{Q^2}$ determining 
 the magnitude of the pion mass corrections is small for $Q^2\geq 2$. 
 Namely, ${m_\pi^2}/{Q^2}< 1 $\% . Thus, the corrections of the
 $\xi$-scaling type \cite{Georgi:1976ve} may be neglected. 
 There remains the contribution of the twist-3 operators 
 $\bar \psi \gamma_5 \stackrel{\leftrightarrow}{D}  \ldots 
    \stackrel{\leftrightarrow}{D} \psi$.
    It can be calculated in the same way as the axial contribution.
 One should also include the next term in the expansion of the function 
 $E(1, \bar g (Q^2))$ having the order of magnitude of $\alpha_s(Q^2)/\pi
 \approx 20$\%. 
 
 Thus, we observe that the agreement 
 of quantum chromodynamics for   $\alpha_s(2) = 0.7$
 with experimental data is not bad, at least in the region
 $1 \leq Q^2 \leq 3 \,{\rm GeV}^2$.

    The parton formulation of our approach can  be easily  applied 
to other hadrons. The proton in  deep elastic scattering
is described by the wave function $\varphi (x_1,x_2,x_3, \mu^2)
\delta (1-x_1-x_2-x_3)$ related to operators of the
$\psi_1 {\cal C}\gamma_5 \gamma_\mu \psi_2 
( \stackrel{\leftrightarrow}{D}_{23})^{N_2}
 ( \stackrel{\leftrightarrow}{D}_{13})^{N_1} \psi_3$ type, 
 where ${\cal C}$ is the charge conjugation matrix. 
 Unfortunately, there is no {\it a priori}
 information about normalization of the function 
 $\varphi (x_1,x_2,x_3, \mu^2)$. 
 It can be extracted from the data on the proton form factor,
 and then the parton wave functions can be used 
 to fit the data on deep elastic wide-angle hadron scattering:
  \begin{eqnarray} 
  M(s,t)|_{AB\to CD} &=& \int \prod dx_k^{(A)} dx_l^{(B)}dx_m^{(C)}dx_n^{(D)}
  \delta \left (1-\sum x_k^{(A)}\right ) \delta \left (1-\sum x_l^{(B)}\right ) 
  \delta \left (1-\sum x_m^{(C)}\right )
 \delta \left (1-\sum x_n^{(D)}\right )
 \nonumber \\
 &\times & \varphi^* (x_m^{(C)})\, \varphi^* (x_n^{(D)})
 E\left ( x_k^{(A)}, x_l^{(B)}, x_m^{(C)}, x_n^{(D)},P_A,P_B,P_C,P_D,\mu^2 \right )
 \varphi (x_k^{(A)})\,\varphi (x_l^{(B)})  \  .
  \end{eqnarray}
 
The  lowest-order approximation for $E$  reproduces the quark counting rules
\cite{Matveev:ra,Brodsky:1973kr} $d\sigma/dt|_{AB \to CD} \sim s^{-N+2}$,
where $N= n_A+n_B+n_C+n_D$. In addition, there appears 
a factor $[\alpha_s (ut/s)]^{N-2}$ that gives an extra
dependence on $s$. In the region $ut/s\sim m_p^2$
one should take into account the proton mass corrections.
The function $Q^2 F_\pi^2 (Q^2)$ behaves like 
$\alpha_s (Q^2)$ even for moderately large $Q^2$.
For this reason, the investigation of the pion electromagnetic 
form factor seems to be the best tool for the
experimental study of the nature of the quark-gluon coupling 
constant renormalization. The wave functions
$\varphi (x, Q^2)$ also depend on $Q^2$ 
due to the anomalous dimensions.
However, preliminary estimates show that
the change of $\gamma (Q^2)$ with $Q^2$ 
cannot compensate the decrease of $\alpha_s (Q^2)$ 
in a non-Abelian asymptotically free
gauge theory, and the absence of the extra 
logarithmic dependence in the region 
$Q^2 \gtrsim m_\rho^2$  would be a serious 
argument against the asymptotic freedom 
$\alpha_s (Q^2) \sim 1/ \ln (Q^2/\Lambda^2)$ 
in favor of the assumption \cite{Efremov:vx} 
that $\alpha_s (Q^2) ={\rm const}$  for $Q^2 \geq 2 \,{\rm GeV}^2$.

\vspace{5mm}

I am deeply grateful to A.V. Efremov for stimulating 
discussions.

\end{document}